# Future Intelligent Data link and Unit-Level Combat System Based on Global Combat Cloud

Ma Xinyan*, Li Wei, Zhong Jian, Li Jinyu, Wang Zheng

China South Industries Group Corporation, Hangzhou Zhiyuan Research Institute Co., Ltd., 310012
China School of Electric Science and Engineering, National University of Defense Technology, P. R. China
Email: { vanessamxy@163.com}

*Abstract*—The development of U.S. Army and NATO's data link systems is introduced first, and then the development trend of future intelligent data link is summarized into integration, generalization, multifunctionality and high security. A unit-level combat system architecture based on the global combat cloud, which is capable of realizing the flexible scheduling of global combat resources and maximizing the overall combat effectiveness, is proposed. Intelligent data link is an important part of this solution, providing strong information support for future urban unit-level warfare.

## I. INTRODUCTION

With the continuous evolution of technology, the nature of war is constantly changing, and the advantage in warfare is shifting rapidly from mechanization towards informatization and intelligence. Future intelligent warfare will possess some new characteristics:

1. Highly informatized and data-driven. Intelligent warfare will rely on highly informatized systems that can collect, process, analyze and utilize large amounts of data in real time, including battlefield environment data and information on the battlefield situation, actions and decisions of the enemy and us. Through big data analysis, it will be possible to more accurately predict and respond to changes on the battlefield and improve operational efficiency.

2. Highly autonomous and intelligent decision-making. Intelligent war will see the emergence of highly autonomous combat units that can perceive, analyze and make decisions on their own without human intervention. At the same time, intelligent war will also rely on artificial intelligence technology, through machine learning and deep learning and other technologies, so that the combat unit can self-optimization and improvement, to improve combat effectiveness.

3. Taking "the right to control intelligence" as the core. In the military field, intelligent warfare has gradually subverted the traditional form, presenting the characteristics of algorithmization of combat command, unmanned combat forces, and diversification of combat styles with the core of seizing the "right to control intelligence". This means that in an intelligent war, acquiring and controlling "the right to control intelligence" will become the key to victory in the war.

4. Mixed games. Intelligent warfare is no longer just a confrontation in the military field, but a hybrid game involving the economy, diplomacy, public opinion, culture and other fields. This makes intelligent war more complex and unpredictable.

Among the many intelligent war features, communication data link is one of the cornerstones. The data link connects the combat elements on the battlefield into an organic whole, and each user transmits data based on a common communication protocol and in accordance with a fixed transmission format, realizes seamless

---
Ma Xinyan is the corresponding author.

data/information linking of each combat element with a unified messaging standard and a common processing protocol, forms a unified battlefield battlefield situation through distributed data processing, and is oriented towards the The tactical tasks are achieved by commanding and controlling the combat elements, realizing the complementary advantages of combat elements and resource sharing, and effectively implementing precision strikes.

Throughout the high-tech wars in recent decades, the data link has played a rather important role, which has profoundly affected the process and the end of the war. The data link runs through the battlefield command center system, weapon and equipment systems, combat units and sub-task command platforms, constituting an all-dimensional integrated information network, sharing all kinds of information and intelligence resources, effectively implementing precision strikes, and is the key and effective means of transforming the information superiority into combat power. It is a key and effective means to transform information superiority into combat power.

Data link realizes efficient data transmission through links, link nodes and link relationships, and weapon and equipment systems equipped with excellent data link can support diversified functions and have the ability of broad combat missions [15]. Efficient, accurate and stable data link makes it have stronger offensive and defensive capabilities, more efficient integrated combat efficiency, more rapid maneuver response capability, and more accurate search and positioning target capability [1].

However, with the rapid development of the Internet of Things, cloud computing, artificial intelligence and other new technologies, intelligent science and technology will permeate the whole process of the whole elements of war, accelerating the evolution of war patterns to intelligence. Battlefield space from the traditional "land, sea, air, sky, electricity, network" physical domain to the ubiquitous social domain, cognitive domain expansion, the Internet of Things has become the basis of war, the physical domain, information domain, cognitive domain, social domain, the depth of integration of the four domains, so that the battlefield holographic transparency, intelligent combat on the data link has also put forward new requirements.

In this paper, we first introduce the development of data link system, then study and summarize the new requirements of the future intelligent tactical data link, and finally propose a new unit-level combat system architecture based on the global combat cloud, which points out the future application of data link.

II. Introduction to U.S. Army and NATO's Data Link Development Systems

Data link research first began in the 1950s, with the continuous development and transformation of operational needs, as well as the rapid development of communications technology and information processing technology, data link performance continues to improve, the data link system is also developing rapidly, with the U.S. Army and NATO as the representative of the developed countries have constructed a more complete data link system architecture, and formed the Link, Common Data Link (CDL), etc. as the representative of the equipment system[2-5]. This section summarizes the development of the data link system of the U.S. Army and NATO.

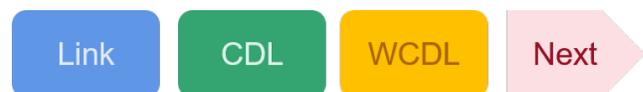

**Fig. 1. U.S. Army and NATO's Data Link Development**

The earliest data link system to be developed was the Link system, Link being the NATO and U.S. military term for the Tactical Data Link, which is the primary data link used for operational command and weapons control systems. Tactical Data Link, also known as Tactical Data Information Link (TADIL), is a link for transmitting tactical data. It has three basic characteristics: first, it is tactical, that is, it is communication between tactical-level users; second, it is data, that is, it is communication in the form of data; and third, it is link, that is, it is communication according to the link protocol.

The U.S. Army and NATO member countries have been developing and using tactical data links since the early 1960s, including more than a dozen types, of which those in common use include Link-4A (TADIL-C), Link-11 (TADIL-A), Link-11B (TADIL-B), Link-16 (TADIL-J), and Link-22 (TADIL-F).

Link-4 is an unclassified datalink used to provide radio guidance commands to fighter aircraft, but Link-4 can only transmit in one direction. The Link-4A and Link-4C were later developed from the Link-4 to support bi-directional transmissions and play an important role in ground-to-air, air-to-ground and air-to-air tactical communications.

Link-11, also known as TADIL-A, is a tactical data link used by the U.S. Navy for the two-way exchange of air track, command and control and other data between ships, between ships and aircraft, between the fleet and the Marine Corps, and between the fleet and land. The main use of high-frequency propagation, in the line-of-sight range also use ultra-high frequency to realize the association of various combat platforms, in the form of a name call, to provide communications and exchange data and information for the various forces.

Link-16 is a high-capacity, classified, jam-resistant tactical data link widely adopted by the U.S. Navy, Air Force, Army, Joint Forces, and NATO forces. Link-16 supports the functionality of Link-11 and Link-4A with added voice, relative navigation, electronic warfare, etc. Link-16 supports both fixed and variable message formats. It has made significant system improvements such as no core nodes, anti-jamming, communication flexibility, secure separation of transmission and data, increased number of participants, increased data capacity, network navigation features, and classified voice [17]. Link 16 is not yet a complete replacement for Link-11 and Link-4A, but it could be considered a better option.

Link-22 is mainly a data link developed by NATO, which is the upgraded version of Link-11, also known as NATO Improved Link-11 [18]. It is a classified, anti-jamming over-the-horizon tactical communication system, mainly used for operational command and tactical coordination between naval shore, ship and ship-air, and can be used between various platforms on land, water, underwater, in the air or in space for the exchange of electronic warfare data as well as the transfer of command and control instructions and intelligence information. Link-22 is compatible with Link-16 and belongs to the wide-area Link-16 family. Link-22 also adopts the TDMA time division multiple access scheme, uses frequency hopping mode in the HF and UHF bands to improve the anti-jamming capability, and provides better interoperability and superior performance through the intelligence automation network management technology. Provides better interoperability and performance through intelligence automated network management technology.

Although the Link series data link can transmit information effectively, the data transmission rate cannot meet the requirements of ISR (Intelligence, Surveillance, Reconnaissance) image transmission. Therefore, the U.S. Department of Defense developed the CDL in the 1980s and named it ISR Standard Data Link in 1991 as a standard link for transmitting images and signal intelligence between satellites, reconnaissance planes and unmanned aerial vehicles (UAVs) and ground control stations [14]. CDL data rates are as high as 274 megabits per second (Mbps) and generally around 10.71 Mbps. Currently, CDL has been fielded on a variety of major ISR platforms, including U2 reconnaissance aircraft and tactical aircraft reconnaissance pods.

The Link series of tactical data links are capable of interoperating with a variety of platforms over the network for a wide range of information, and the links are flexible and can meet a variety of needs, but the bandwidth is narrower; the CDL, on the other hand, utilizes a series of interoperable data links, which can be expanded and configured in a variety of ways to meet the use of special application platforms. Configuration options include, for example, the frequency band used (X or Ku band), the data rate (up to 274 Mbit/s), and the transmission power. Interoperability between the CDL series of data links can be achieved by specifying the data link waveforms (RF and digital), controlling and coordinating hardware configuration options, and so on [12].

Another type of new data link that deserves attention is the Weapon Coordination Data Link (WCDL), which develops precision-guided data links for weapon combat platforms to support sensors in the theater of operations to be connected into a network, share sensor data in real time, and generate a single comprehensive image of the theater of operations. Taking the application of WUCC in flight combat as an example, due to the use of WUCC for high-speed wireless communication, aircraft crews can realize real-time and reliable information sharing and enhance the crew's perception of the airspace, so that the flight force can strategically attack the enemy in a larger range [6]. At present, such data links mainly include Cooperative Engagement CapabilityTechnology (CECT), Airborne Data Link Technology (ADLT), Tactical Targeting Network Technology (TTN) and Tactical Targeting Network Technology (TTN). Targeting Network Technology (TTNT) and Quint Networking Technology (QNT) [1]. TTNT, for example, is a solution with high transmission capacity and short response time to solve the problem of data link from "sensor to shooter". Based on the Internet Protocol (IP), it is a high-speed, dynamic, dedicated network that enables the U.S. military to rapidly target moving and time-sensitive targets [9-11].

### III. Reflections on the Future Development of the Data Link System

With the change of war pattern, the requirements for data link in the future intelligent battlefield are getting higher and higher. In this section, we put forward several reflections on the future development trend of data link system.

*A. Integrated Data Link*

With the development of 6G technology, the development of air-heaven-air-sea integrated networking technology potentially drives the development of future data links to extend in the direction of air-heaven-air-sea integration as well. In the future intelligent battlefield, the integration of air, sky, earth and sea is an inevitable trend of development, which can provide wide-area and deep coverage within the scope of tactical activities, and provide various types of combat units and nodes with on-demand tactical services, secure access and on-demand tactical services, and support the integration of "detecting, controlling, fighting, and evaluating" and the all-area combat [2][16].

Currently, the U.S. military is also extending and expanding the use of data links. For example, the Space CDL is a technical layout of the U.S. Army to extend the CDL from space-based to space-based. Space CDL has successfully verified the tactical application capability of imaging reconnaissance satellites and will become an important means for the U.S. military to build a multi-source battlefield intelligence system. The space-based universal data link uses satellite channels to realize the universal data link format transmission data link, which is mainly used for the transmission of satellite and ground reconnaissance intelligence surveillance data, with full duplex, anti-jamming and other characteristics, and the satellites used include military and civilian communications satellites [13].

With the development of data link systems and related technologies, space-based data link, ground-based data link, sea-based data link and so on are fully developed respectively, the data link networks in various fields will certainly be developed like fusion and integration, constituting an omni-domain fusion ubiquitous data link network, which will jointly support the future network information system.

*B. Generic Data Link*

Another major future trend in data links is the shift from specialized to generalized. Generalized datalinks are generic communications systems that can be used for a wide range of purposes and platforms. They have the flexibility to adapt to diverse mission requirements in different battlefield environments. Universal datalinks can support a variety of communications and messaging missions, including voice, data, image, and video transmission. This versatility improves the interoperability and efficiency of the military.

Early data links were basically developed independently by each military service according to their respective operational needs, but with the development of informationized combat, the future battlefield form is bound to be multi-military and military services integrated joint operations, and data links equipped on different military service platforms need to be interconnected with the network and interoperable with each other in order to realize the joint command and operation. Therefore, the data link in the future intelligent battlefield will be developed from specialization to generalization, and the interoperability of the data link will be significantly improved.

In order to achieve the ability to effectively share information and communicate between different military platforms and equipment without the need for large-scale customized integrations or modifications, a number of specific technical means are required. For example, common communication protocols and standards need to be adopted by different military platforms and equipment so that they can understand and interpret the information they send to each other. Information in the data link also needs to be transmitted according to a uniform data format and encoding, which ensures that the information can be correctly decoded and processed at the receiving end. Accordingly, soldiers and operators need to be uniformly trained in the proper use and operation of the different datalink devices. Ultimately, increased interoperability in diverse battlefield environments will be realized, leading to improved operational effectiveness and decision-making capabilities.

Of course, specialized datalinks will still play their unique advantages in specific application scenarios, because they usually have higher performance and security.

*C. Multifunctional Software-enabled Data Link*

With the development of intelligent technology and equipment, the requirements for data links are more complex, requiring the transmission of different information contents in different links, and there are great differences in transmission bandwidth and objects. At the same time, the space and load resources of air platform are tight, and the independent retrofitting and transformation of each type of datalink will cause the accumulation of the number and weight of equipments, which will cause great pressure on the platform space layout, payload, power consumption of equipments, electromagnetic compatibility and even the cost of equipments [7]. Therefore, the development of future data link system will be transformed to multifunctional and software-based.

The use of software radio terminals to realize the multifunctional software of data links is one of the possible development directions. At present, there are many models of data links and different hardware platforms, if software radio technology can be used to design the data link system, it can realize the unity of the hardware platform, the flexible configuration of the system functions, the new functions and new technologies can be realized through the timely application of software upgrades, which is convenient for the system to upgrade and maintain, and greatly shortens the development cycle. However, the data link software radio technology also exists some key issues that need to be resolved, such as the requirement of software development modularity, flexible software development technology and hardware platform with high performance [8][17].

The U.S. Army has already engaged in a software-based process for data links, using the Software Communications Architecture (SCA) developed by JTRS as the standard software architecture for embedded systems to provide a standard, open and interoperable radio communications software platform to ensure that all types of data link systems can be loaded and unloaded on the platform in a modular and software-based format.

*D. High Security and High Reliability Data Link*

Another important need for datalinks in the future war is in the area of security. Security is a fundamental prerequisite for military communications and information transfer, and as cyber threats increase, future tactical datalinks will need to employ stronger encryption and cybersecurity measures to prevent hostile forces from intruding or interfering with communications. The data link, as a communication link for transmitting important operational information such as charge information and ammunition information, must ensure security and reliability. Consideration can be given to utilizing blockchain technology and endogenous security technology, etc., on the one hand, to realize the confidential transmission of system information and complete the anti-interception transmission, and on the other hand, to carry out research on strong anti-destructive and anti-jamming technology in terms of waveform selection, coding method, and network topology adjustment.

IV. Operational Cloud-based Unit Level System Architecture

Based on the reflections above, we focus on urban unit combat scenarios and propose a unit-level system architecture based on global combat cloud supported by data link, and the overall structure is shown in Figure 2.

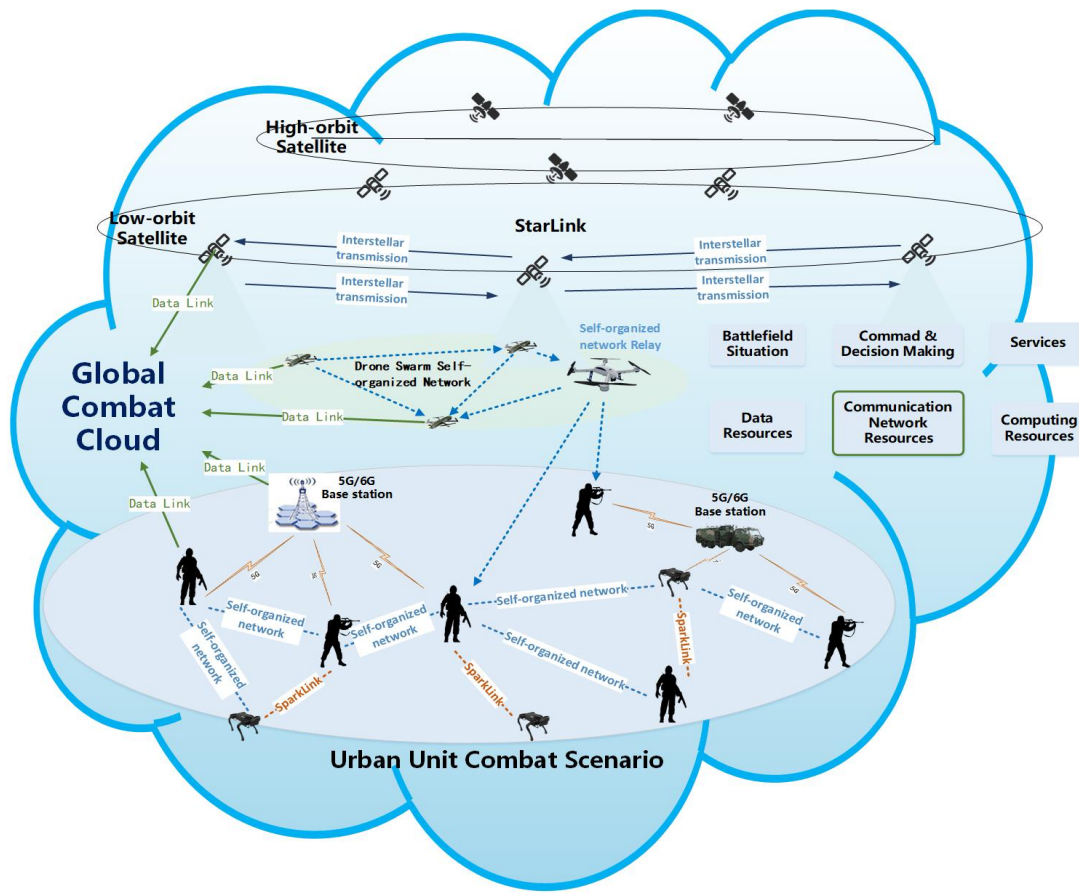

**Fig. 1. Unit-level system architecture based on combat cloud**

The whole architecture is in a typical unit-level urban combat scenario, consisting of ground-based unmanned cooperative units, various types of unmanned aircraft, star chains composed of airborne high- and low-orbit satellites, and a global combat cloud. Within the unit, each combat unit builds a unit network through self-organized network, existing 5G/6G base station, star chain and other ways, with different network chains complementing each other's strengths, redundancy and backup, and information interoperability, to jointly build a unit-level communication and interconnection network.

Cloud is characterized by resource virtualization, elastic management, and flexible sharing, and the organic aggregation of combat resources into a combat resource pool becomes a combat cloud. The global combat cloud in this architecture is a kind of cloud resource pool, which contains all kinds of battlefield information, combat resources and services. More specifically, the global combat cloud contains the global situation information of the whole battlefield, the real-time decision-making information completed by the commander, communication network resources, arithmetic resources, basic data resources, and all kinds of services that can be invoked, and so on.

In the architecture, each combat unit is a part of the combat cloud, uploading its own combat resources (including weapon and equipment resources, battlefield situational awareness, etc.) to the combat cloud to constitute a resource pool, and at the same time also being able to understand in real time the real-time situation of the other areas of the battlefield and the available resources through the combat cloud, so as to achieve global omniscience of the battlefield situation. Further, each combat unit is able to invoke required services, such as fire strike support, higher communication bandwidth resources, etc., through the combat cloud. Therefore, based on the architecture of the combat cloud, each combat unit can independently fulfill one or more of the functions of reconnaissance, command, and strike, and ultimately enable group intelligence to achieve combat objectives.

In order to realize the above architecture, a number of key technologies are required. For example, the design of the overall combat management system, the construction of the communication network within the architecture, and the method of integrating the combat capabilities of different military services. Among them, the communication system, as the communication guarantee and information base of the overall architecture, is an indispensable condition for the operational effectiveness of the architecture.

Intelligent data link, as an integral part of the communication system, plays a very important role in the above combat cloud system. The transmission of charge information, ammunition information, strike results and other combat resource information between all combat units and the combat cloud will be accomplished through the data link. Therefore, it is necessary for the data link system to conform to the trend and demand of the development of future intelligent war, and to form an air-ground fusion, function-rich, safe and reliable communication link.

## V.   CONCLUSION

The development history and research status of U.S. Army and NATO's data link system is introduced first, and then the development direction of future intelligent data link is analyzed and envisioned. The development trend is summarized into integration, generalization, multifunctionality as well as high security. Finally, a unit-level system architecture based on the global combat cloud is proposed. This scheme can realize the flexible scheduling of global combat resources, so that each combat unit can grasp richer resources such as battlefield situation, firepower, and computing power, and thus be able to complete one or more tasks in reconnaissance, command, and strike independently, maximizing the overall combat effectiveness. Intelligent data link is an important part of the scheme, providing strong information support for the future urban unit-level warfare.